\documentclass[12pt]{iopart}
\usepackage{amsfonts,amsthm,latexsym,amssymb}
\usepackage{graphicx,lscape,fancyhdr,array}

\newcommand{\p}[1]{(\ref{#1})}

\newcommand{\be}{\begin{equation}}
\newcommand{\bea}{\begin{eqnarray}}
\newcommand{\ee}{\end{equation}}
\newcommand{\eea}{\end{eqnarray}}

\def\dsll{\not {\! \partial}}

\def\a{\alpha}

\def\d{\delta}

\begin{document}

\title{On Higher Spin Theory:  \\ Strings, BRST, Dimensional Reductions }

\author{X.~Bekaert\dag,
  I.L.~Buchbinder\ddag,
 A.~Pashnev\S \\
and M.~Tsulaia\dag\S *}

\address{\dag Dipartimento di Fisica, Universit\`a degli
Studi di Padova, \\ INFN, Sezione di Padova\\Via F. Marzolo 8,
35131 Padova, Italy}
\address{\ddag Department of Theoretical Physics, Tomsk State
Pedagogical University, \\ Tomsk, 634041, Russia}
\address{\S Bogoliubov Laboratory of Theoretical Physics, JINR, Dubna, 141980
Russia}
\address{*Institute of  Physics, GAS \  380077 Tbilisi \
Georgia}

\begin{abstract}
We briefly review some modern developments in higher spin field
theory and their links with superstring theory. The analysis is
based on various BRST constructions allowing to derive the
Lagrangians for massive and massless higher spin  fields on flat
or constant curvature backgrounds of arbitrary dimensions.
\end{abstract}

%Uncomment for PACS numbers title message
%\pacs{00.00, 20.00, 42.10}

% Uncomment for Submitted to journal title message
%\submitto{\JPA}

% Comment out if separate title page not required
%\maketitle

\section{Introduction}

The paper under consideration is devoted to a brief review of a
modern approach to the description of higher spin (HS) fields
propagating on flat and constant curvature backgrounds. This
approach is based on the BRST technique and has many common
features with string field theory. Such a link between HS field
theory and (super)string theory is obviously not just formal since
the theory of HS fields \cite{V} is conjectured to be an
interesting -- though not thoroughly investigated -- corner of
modern string and M--theory.

Being first introduced as a tool for the quantization of gauge
theories, the BRST formalism then turned out to be very powerful
for the formulation of field theories which correspond to given
first quantized systems. A famous example is string field theory
\cite{sft}  that proved to be very useful for the study of string
interactions and various superstring vacua containing
nonperturbative objects such as D-branes.

After reviewing the triplet structure of HS fields and the way it
arises in the tensionless limit of open bosonic string theory
\cite{FS} -- \cite{Sagnotti:2003} we shall discuss various
relevant BRST constructions \cite{Mirian} -- \cite{BuPT} and
present a general method for constructing HS massive theories in
flat and constant curvature backgrounds. Finally, we shortly
describe possible further developments such as the supersymmetry
of triplet equations and other open problems.

\section{Triplets and tensionless limit of string theory}\label{tripletsec}

To start with, let us introduce the Fock space spanned by the
annihilation and creation operators $\a_1^M$ and
$\a_{-1}^M:=(\a_1^M)^+$ satisfying $[\a^M_1 , \a_{-1}^N] = g^{MN}$
where $g_{MN}$ is a flat space metric with the mostly plus
signature. A completely symmetric tensor field of rank $s$ can be
represented in this Fock space by the vector
\begin{equation} \label{Fockvector}
|\,\varphi\rangle = \frac{1}{s!}\,\varphi_{M_1 M_2\cdots M_{s}}
(x)\, \a_{-1}^{M_1}\a_{-1}^{M_2 }\cdots \a_{-1}^{M_s}
|\,0\rangle.\nonumber
\end{equation}
The physical field $|\,\varphi \rangle$ must satisfy the
mass-shell, the transversality, and the tracelessness constraints,
i.e it should be annihilated by the operators
\begin{equation} \label{lapl}
\tilde L_0 = g^{M N}(P_M P_N + i\,\Gamma^L_{M N}P_L)= P^A P_A -
i\, \Omega_A{}^{A B} P_B
\end{equation}
\begin{equation} \label{tr}
\tilde L_1 = P_M \a_1^M, \quad \tilde L_{11}=\frac{1}{2}\, g_{MN}
\a_1^M \a_1^N,
\end{equation}
where the operator $P_M$ is the covariant derivative $P_M := - i
(\partial_M + \Omega_M{}^{A B}\a_{-1\,A} \a_{1\,B})$ while
$\Omega_M{}^{A B}$ is the spin connection. Let us note that,
although in this section we will be dealing with a flat space, the
operators (\ref{lapl}) -- (\ref{tr}) are presented in the generic
form since this form will be used in curved backgrounds.

As is well known, in the standard formulation of HS gauge fields
\cite{Fronsdal} the basic object is a doubly traceless completely
symmetric tensor field. This can be provided by the off-shell
constraint \footnote{The rule to make contact with the notation
used in the works \cite{FS,Sagnotti:2003,Francia:2002} is to make
the replacements: $|\,\varphi\rangle \rightarrow \varphi$,
${\tilde L}_0\rightarrow -\Box$, ${\tilde L}_1\rightarrow
-i\partial\,\cdot\,$, ${\tilde L}_{-1}\rightarrow -i\partial$,
$2{\tilde L}_{11}\rightarrow {}^\prime\,$ and $F_0\rightarrow
-i\dsll\,$.} \be ({\tilde L}_{11})^2|\, \varphi\rangle
=0\label{doubletr}\ee which is invariant under the gauge
transformations $\delta |\varphi \rangle = {\tilde L}_{-1}
|\,\Lambda \rangle$ if the parameter of gauge transformations is
restricted to be traceless, ${\tilde
L}_{11}|\,\Lambda\rangle=0\,$. We defined ${\tilde
L}_{-1}:=({\tilde L}_1)^+\,$. In order to avoid the off-shell
constraints on the gauge transformation parameter and basic field,
an alternative formulation of higher spin theory was suggested
\cite{Francia:2002} where the field equations are nonlocal
(tensors of mixed symmetry were recently discussed in these terms
in \cite{mixednloc}). One way to obtain the nonlocal equations is
via the triplet structure of HS fields \cite{FS} though the
triplet provides an ``unconstrained" description of the HS fields
by itself. More specifically the triplet consists of the
completely symmetric fields $\varphi$ (of rank $s$), $C$ (of rank
$s-1$) and $D$ (of rank $s-2$) satisfying \be
 {\tilde L}_0  |\,\varphi\rangle\,  =  {\tilde L}_{-1} |\, C\rangle \,,
 \quad |\, C\rangle  =  {\tilde L}_1|\, \varphi\rangle  -
{\tilde L}_{-1} |\, D\rangle \,,\quad {\tilde L}_0 |\, D\rangle\,
=  {\tilde L}_1|\, C\rangle \,,\label{triplet}  \ee and the gauge
transformation rules
\begin{equation}
\delta |\,\varphi\rangle = {\tilde L}_{-1} |\,\Lambda\rangle,
\quad \delta|\, C\rangle = {\tilde L}_0|\, \Lambda\rangle, \quad
\delta |\,D\rangle = {\tilde L}_1 |\,
\Lambda\rangle\,.\label{grules}
\end{equation}
One can conclude that a triplet, after the suitable gauge fixing,
describe the propagation of fields of spins $s,s-2,s-4,\ldots\,$
In order to describe the irreducible representations, an
additional field -- the compensator $|\,\alpha\rangle$ -- should
be introduced. It transforms via the trace of the gauge
transformation parameter $\delta|\, \alpha\rangle = {\tilde
L}_{11}|\,\Lambda\rangle$ and obeys the equation
\begin{equation}
{\tilde L}_{11}|\,\varphi\rangle -|\, D\rangle ={\tilde L}_{-1}
|\,\alpha\rangle\label{compensator}
\end{equation}
Such a compensator allows to describe a single irreducible higher
spin field.

The triplet structure described above finds its natural origin in
string theory \cite{FS}. For instance one can take the BRST charge
for bosonic open string \footnote{We follow the notations of
\cite{book}.}, make the redefinition of $bc$ ghosts as $c_m
\rightarrow \sqrt{2 \alpha^\prime} c_m$, $b_m \rightarrow
\frac{1}{ \sqrt{2 \alpha^\prime}} b_m$, for $m \neq 0$
 and $c_0 \rightarrow  \alpha^\prime c_0$, $b_0
\rightarrow \frac{1}{ \alpha^\prime} b_0$, and then go to the
$\alpha^\prime \rightarrow \infty $ limit. Alternatively, one can
take the tensionless limit in the Virasoro constraints prior to
the BRST construction \cite{Henneaux}. The resulting BRST charge
is\begin{equation} Q =c_0  {\tilde L}_0 +
\sum\limits_{m=1}^{\infty} (c_{-m} {\tilde L}_m + c_{m} {\tilde
L}_m^+ +c_{-m}\,c_{m}\, b_{0})
\end{equation}where the redefined Virasoro generators satisfy the algebra $[{\tilde L}_m,{\tilde L}^+_n]=\d_{mn}{\tilde L}_0$.
If we restrict the string field to the most general level $s$
leading Regge trajectory \be |\, \Phi \rangle =  |\, \varphi
\rangle \ + c_{-1}\,b_{-1}|\, D\rangle + \ c_0 \,b_{-1} |\,\, C
\rangle\ee and insert it into the BRST cocycle condition
$Q|\,\Phi\rangle=0$, then we get the triplet equations
(\ref{triplet}). The string field gauge transformation
$\d|\,\Phi\rangle=Q|\,\Sigma\rangle$ corresponds to the
transformation rules (\ref{grules}) if one takes
$|\,\Sigma\rangle=b_{-1}|\,\Lambda\rangle$.  One can consider the
general dependence of the string field on an arbitrary finite
number of oscillators and recover the generalization of
(\ref{triplet}), i.e. the generalized triplet \cite{Sagnotti:2003}
which exhausts the bosonic string spectrum in the tensionless
limit. Let us also mention that the triplet structure can be
successfully deformed to the case of an arbitrary dimensional
(A)dS space as well \cite{Sagnotti:2003}.

\section{The dimensional reduction}\label{reduction}

Dimensional reduction might be the most elegant way to produce
massive field theories from their known massless counterparts. In
this way the massive triplet is recovered, which -- besides of
having a different field content compared to the usual tensile
string equations -- is gauge invariant in any dimension rather
than only for $D=26$. The detailed connection of the massive
generalized triplet equations and string mass generation mechanism
is a still challenging open problem \cite{Sagnotti:2003}. Here, we
concentrate on another side of the dimensional reduction.

It is known that HS fields propagate consistently in flat and
constant curvature backgrounds, all of which can be sliced in
codimension one constant curvature spaces. Among all corresponding
possibilities, three types of dimensional reductions were
considered in the literature on higher spins: (i) the usual road
going from ${\mathbb R}^{D,1}$ to ${\mathbb R}^{D-1,1}$
\cite{Aragone,P}, (ii) the radial reduction from ${\mathbb
R}^{D,1}$ (${\mathbb R}^{D-1,2}$) to $dS_D$ ($AdS_D$) recently
proposed in \cite{BS}, and (iii) the reduction from $AdS_{D+1}$ to
$AdS_D$ \cite{M}. We will only discuss the first two types of
reductions in the BRST framework, nevertheless the procedure we
present is completely general (in the sense that one could start
the reduction from any $D+1$ dimensional  flat or (A)dS
background).

The simplest way to make the dimensional reduction and thus
generate the masses for triplets is to take the Lagrangian
\begin{equation} \label{L}
{\cal L} =  \langle \Phi|\, Q |\, \Phi \rangle,
\end{equation}
and make $x^D$ coordinate compact,
 i.e. take  the following ansatz
for the string field dependence on $x^D$ \cite{P,PT1}
\begin{equation} \label{vector}
|\,\Phi \rangle = V |\,\Phi^\prime \rangle\,,
\end{equation}
where $V= e^{imx^D}$. This is equivalent to the unitary
transformation of the BRST charge
\begin{equation} \label{DR}
Q \rightarrow V^+  Q V,
\end{equation}
Obviously the resulting BRST charge is also nilpotent and
hermitian in any dimension.

Being the simplest possible one, (\ref{DR}) gives a hint on how to
deal with more complicated cases of second class constraints.
Indeed, from the point of view of ${\mathbb R}^{D-1,1}$, the
operators $L_{\pm 1}$ and $L_0\equiv P^2+m^2$ form second class
constraints\footnote{Operators with tilde will now correspond to
the higher dimensional flat space ${\mathbb R}^{D,1}$, while
operators without tilde correspond to ${\mathbb R}^{D-1,1}$ or
$(A)dS_D$ space ($M=0,1,\ldots,D$ and $\mu=0,1,\ldots,D-1$).}. A
general way to deal with such a problem is therefore to go to a
larger phase space where the standard BRST technique applies
\cite{BPT}. In the present illustrative case, we go one dimension
up and consider the operators $\tilde{L}_{\pm 1} = P_\mu\,
\alpha_{\pm 1}^\mu + P_D\, \alpha_{\pm 1}^D\,$, $\tilde{L}_0 = P^2
+ P_D^2$ which form first class constraints.

Another kind of reduction is needed when the resulting $D$
dimensional space is  curved, since in that case the
Kaluza-Klein-like reduction discussed above does not eliminate the
dependence on the extra coordinate in the $D$ dimensional
Lagrangian. Let us consider the procedure of dimensional reduction
from flat spaces in its full generality. We initially include the
operators $\tilde L_{11}$ and $\tilde L^+_{11}$ into the set of
constraints because they are known to be unavoidable in $(A)dS_D$
space. In the flat $D+1$ dimensional space with signature
$(-+++...+\kappa), \kappa=\pm 1$, the constraints -- together with
the operator ${\tilde G}_0 =g_{MN}\,\a_1^{+M}\a_1^N +
\frac{D+1}{2}\,$ -- form the following algebra
\begin{eqnarray} \label{alf}
&&[\tilde L^+_1, \tilde L_1] = -\tilde L_0,  \quad [\tilde
L^+_1,\tilde L_{11}] = -\tilde L_1, \quad
[\tilde L^+_{11},\tilde L_1] = -\tilde L^+_1, \nonumber \\
&&[{\tilde G}_0, \tilde L_1] = -\tilde L_1, \quad [\tilde
L^+_1,{\tilde G}_0] = -\tilde L^+_1,
\end{eqnarray}
with the $SO(2,1)$ subalgebra
\begin{equation} \label{so21f}
[{\tilde G}_0,\tilde L_{11}] = -2 \tilde L_{11}, \quad [\tilde
L^+_{11}, {\tilde G}_0] = -2\tilde L^+_{11}, \quad [\tilde
L^+_{11}, \tilde L_{11}] = -{\tilde G}_0.
\end{equation}Hermitian conjugation is defined with respect to
the integration measure $\sqrt{-\kappa g}$, where $g=det(g_{MN})$.

Next, we construct the nilpotent BRST charge for this system (see
\cite{Sagnotti:2003} -- \cite{BuPT} for the details of the
construction). The main problem is the presence of the operator
${\tilde G}_0$ which is strictly positive and therefore cannot
annihilate the physical state, i.e. one effectively deals with
second class constraints. One can deal with the problem in
complete analogy with the  case of triplet dimensional reduction.
The role of the oscillator $\a_1^D$ will be played by an
additional oscillator $d$. First, we build the auxiliary
realization of $SO(2,1)$ subalgebra in terms of the oscillator $d$
and a constant parameter $h$ provided the auxiliary realization of
${\tilde G}_0$ depends on this parameter linearly. Second, we
define the operators $\hat G_0 :={\tilde G}_0 +{\tilde G}_{0\,aux}
+ h$ and $ \hat L_{11} :={\tilde L}_{11}  + {\tilde
L}_{11\,aux}(h) $ as the sum of the initial and auxiliary
realizations, and third, we construct the standard BRST charge
treating {\it all} operators as the set of first class constraints
\begin{eqnarray}
& Q = c_0 \tilde L_0 + c_1 \tilde L_1^+ + c_{11} \hat L_{11}^+ +
c_1^+\tilde L_1 + c_{11}^+ \hat L_{11}-
    c_1^+c_1b_0 -b_1^+c_1^+c_{11}
 -c_{11}^+ c_1 b_1& \nonumber \label{br}\\
&  - c_G(c_1^+b_1+b_1^+c_1+ 2c_{11}^+b_{11} + 2b_{11}^+c_{11}+\hat
G_0-3)  - c_{11}^+b_G c_{11} .&
\end{eqnarray}
Finally, we consider an auxiliary phase space $(x_h, h)$ and make
the similarity transformation
\begin{equation} \label{DR1}
Q \rightarrow U^{-1}  Q U, \quad U= e^{i\pi x_h}
\end{equation}
where $\pi = -(\tilde G_0 + 2d^+d -3+ c^+_1  b_1  + b^+_1 c_1+2
b^+_{11} c_{11}+2 c^+_{11}   b_{11})$. Thus, one gets rid of the
$c_G$ dependence in the BRST charge while preserving its
nilpotency at the same time. Obviously, after the similarity
transformation (\ref{DR1}) the $b_G$ dependence of $Q$ can be
dropped without affecting the nilpotency of the BRST charge.

Now it is possible to eliminate all $d^+$ dependence in
$|\,\Phi\rangle$ and arrive to triplet and compensator equations
(\ref{triplet}) and (\ref{compensator}) via a partial BRST gauge
fixing \cite{Sagnotti:2003}. If we further gauge away the
compensator $|\,\a\rangle$ and eliminate $|\,C\rangle$ and
$|\,D\rangle$ via their own equations of motion \cite{Mirian},
then we obtain the Lagrangian for completely symmetric massless HS
fields in arbitrary dimensional flat space ${\mathbb R}^{D,1}$
which generalizes Fronsdal's one \cite{Fronsdal} \be{\mathcal
L}_{D+1}=\frac12 \langle\varphi|\,\, K \mid\varphi\rangle\,.\ee
The kinetic operator $K$ is given by \be K = -{\tilde L}_0+
{\tilde L}^+_{1}{\tilde L}_1-({\tilde L}^+_{1})^2{\tilde L}_{11}
-{\tilde L}_{11}^+({\tilde L}_1)^2 +2{\tilde L}_0{\tilde
L}_{11}^+{\tilde L}_{11}+{\tilde L}_{11}^+{\tilde L}^+_{1}{\tilde
L}_1 {\tilde L}_{11}\,. \ee By solving the $D+1$ dimensional
double tracelessness constraint (\ref{doubletr}), we can find the
explicit form for the dependence of $|\,\varphi\rangle$ on the
operators $a_D^+$ \cite{P}. The rank-$s$ field $|\,\varphi\rangle$
is then given in terms of four $D$ dimensional fields
$|\,\varphi_i\rangle$ ($i=0,1,2,3$) of respective ranks $s-i$. The
fields $|\,\varphi_1\rangle$ and $|\, \varphi_2\rangle$ can be
eliminated by fixing the gauge completely.

To address case (i) let us take the dimensional reduction of
lagrangian ${\cal L}_{D+1}$ to the flat space ${\mathbb
R}^{D-1,1}$. To make a connection with the work of Singh and Hagen
in four dimensions \cite{Singh}, one should present
$|\,\varphi_0\rangle$ and $|\,\varphi_3\rangle$ as
\be|\,\varphi_i\rangle=\sum\limits_{m=0}^\infty
\frac{(L_{11}^+)^m}{m!}|\,\varphi_i^{(m)}\rangle\,,\quad
(i=0,3)\,,\label{repl} \ee where the fields $|\,
\varphi_i^{(m)}\rangle$ are all traceless, i.e. $L_{11}|\,
\varphi_i^{(m)}\rangle=0$. This set of fields should correspond to
the minimal set of auxiliary fields. The standard procedure to
formulate free massive HS field Lagrangians \cite{Singh} uses the
traceless fields of spin $s$, $s-2$, $s-3$, $s-4$, ... In the
approach under consideration, one gets the fields
$|\,\varphi_0^{(0)}\rangle$, $|\,\varphi_0^{(1)}\rangle$,
$|\,\varphi_0^{(2)}\rangle$, ... of respective ranks $s$, $s-2$,
$s-4$, ... and the fields $|\, \varphi_3^{(0)}\rangle$,
$|\,\varphi_3^{(1)}\rangle$, ... of respective ranks $s-3$, $s-5$,
... which exactly corresponds to \cite{Singh} at $D=4$. As a
result, we get a possibility to construct a Lagrangian formulation
for massive HS field theory in arbitrary dimensional space. The
previous procedure should clarify the structure of massive HS
field Lagrangian in arbitrary $D$ dimensions.

To address case (ii) let us perform a dimensional reduction of the
BRST action (\ref{L}) where the BRST charge is given by
(\ref{br}). Correspondingly, we now focus on the BRST
generalization (including ghosts) of the radial dimensional
reduction proposed in \cite{BS}. In particular we choose the
parameterization of the $D+1$ dimensional flat space in the form
\begin{equation} \label{sig}
E^A_M=\left( \begin{array}{c c} r\,e^\alpha_\mu & 0 \\ 0 & 1
\end{array}
\right) \,,\quad E^M_A=\left( \begin{array}{c c}
\frac{1}{r}\,e^\mu_\alpha & 0 \\ 0 & 1 \end{array} \right)\,.
\end{equation}
The corresponding tangent space metric is
\begin{equation} \label{sig1}
g_{AB}=\left( \begin{array}{c c} \eta_{\alpha \beta} & 0 \\ 0 &
\kappa \end{array} \right)\,.
\end{equation}
For the case of the $dS_D$ space embedded into the flat space
${\mathbb R}^{D,1}$ one should take $\kappa=1$, while for the case
of the $AdS_D$ space one should take $\kappa=-1$. The vielbein
$e^\alpha_\mu$ and its inverse $e_\alpha^\mu$ describe in each
case the corresponding $D$ dimensional spaces. The measure density
of the flat $D+1$ dimensional space is expressed in terms of the
measure density of the $D$ dimensional space as
\begin{equation}\label{Measure}
\sqrt{-\kappa \,det(g_{MN})}=r^D \sqrt{-\kappa\, det(g_{\mu\nu})}.
\end{equation}
The parameterization \p{sig} gives the simple relations between
the spin connections:
\begin{equation}
\Omega_\mu{}^{\alpha}{}_{\beta} =
\omega_\mu{}^{\alpha}{}_{\beta}\,, \quad \Omega_\mu{}^{\alpha}{}_r
= e_\mu^\alpha\,,
\end{equation}
the other components of $\Omega_M{}^{A B}$ being zero. Therefore
the momentum operator $P_M$ is decomposed as
\begin{equation}
P_\mu = -i(\partial_\mu + \omega_\mu{}^{\alpha\beta}
 \a_{-1\, \a} \a_{1\,\beta})
 + i\,e^\alpha_\mu (\a_{-1\,\alpha}  \a_1^r
- \a_{-1}^r \a_{1\,\alpha})\,, \,\, P_r = -i \,\partial_r\,.
\end{equation}

In analogy with the triplet dimensional reduction, let us fix the
following $r$ dependence in the state vector by inserting in
(\ref{DR}) the following expression for $V$:\be V =
r^{-(\frac{D+1}{2}+iM+N_0+N_1)}\,,\ee where $M$ is a real
parameter related to the mass, and the operators $N_0:= c_0
b_0-b_0 c_0$ and $N_1:=c_1^+b_1-b^+_1c_1$ are ghost number
operators. Note that the transformation (\ref{DR}) is no more
unitary. However the resulting BRST charge $Q$ is nilpotent and
hermitian with respect to the $D$ dimensional integration measure.
The important point is that after the transformation \p{DR} the
Lagrangian \p{L} multiplied by the integration measure
$drd^Dx\times$\p{Measure} becomes independent from the coordinate
$u=\log r\,$. In other words, the proper dimensional reduction has
been done. It could be mentioned that the requirement of
hermiticity and independence on $u$ of the resulting $D$
dimensional Lagrangian completely fixes the explicit form of the
operator $V$.

\section{Conclusions and perspectives}

In this short review we have discussed the connection of massless
HS fields with bosonic string spectrum, and the mass generation
mechanisms for HS fields via various types of BRST dimensional
reductions. Let us note that these results can be extended to the
case of the superstring, say type I. For instance, the fermionic
triplet could be obtained \cite{Sagnotti:2003} after the
truncation of the string functional to the form
\begin{equation} \label{truncated}
|\Phi^R \rangle \ = \ |\Phi_1^R\rangle \ + \ (\gamma_0 \ + \ 2 \,
c_0 \, F_0\,)\, | \Phi^R_2\rangle \ ,
\end{equation}
where $\gamma_0$ is bosonic ghost zero mode and $F_0 = P_\mu
\psi^\mu_0$. Again, we restrict the functionals $|\Phi_1^R\rangle
$ and $|\Phi_2^R\rangle $ to be dependent on only $bc$ ghosts and
$\alpha_1$ oscillators. The mechanism of mass generations could be
reproduced following the lines of the bosonic systems. Moreover,
supersymmetry of triplets can also be established by making use of
the operator $W$ \cite{KNNW1} -- the fermionic vertex emission
operator in CFT language \cite{fms} -- converting the BRST charge
in the R sector to the one in the NS sector
\begin{equation} \label{KOM}
Q_R \, W \ = \ W \, Q_{NS}\,,
\end{equation}
and by avoiding some subtleties -- concerning the closure of
supersymmetry algebra in NS sector \cite{EN} caused by the
presence of pictures -- thanks to the restriction on the
oscillator dependence of the string field (which is possible in
the tensionless limit). It seems also interesting to obtain the
explicit expressions for massive HS fields in flat and constant
curvature backgrounds, and study in detail the spectrum of these
models, thus making an appropriate generalization of \cite{Singh}
that seems relevant in the light of modern studies of massive HS
fields \cite{BS} -- \cite{M}, \cite{BGP} -- \cite{dM}. Another
aspect of massive HS field theory is the problem of constructing
supersymmetric HS models in flat and (A)dS spaces (the first
attempts in this direction have recently been undertaken in
\cite{Bu}).

\section*{Acknowledgments}

We acknowledge  G. Barnich, G. Bonelli, M. Grigoriev, P. Pasti, D.
Sorokin, M. Tonin, M. Vasiliev and especially A. Sagnotti for
stimulating discussions. X.B. thanks the organizers of the RTN
Workshop in Copenhagen for giving him the opportunity to present
this talk. X.B. and M.T. are supported in part by INFN, the EC
contract HPRN-CT-2000-00131. I.L.B., A.P. and M.T. are supported
by  INTAS grant, project No 00-00254. I.L.B. is  grateful to RFBR
grant, project No 03-02-16193 and grant for support of Leading
Russian Scientific Schools, project No 1252.2003.2. A.P. is also
grateful to RFBR grant, project No 03-02-17440 and a grant of the
Heisenberg-Landau program.

\end{document}